\documentclass{iopart}
\usepackage{iopams}
\usepackage{hyperref}
\usepackage{color}
\usepackage{graphicx}

\begin{document}

\title{Using materials for quasiparticle engineering}

\author{G. Catelani$^{1,2}$ and J. P. Pekola$^{3,4}$}

\address{$^1$JARA Institute for Quantum Information (PGI-11), Forschungszentrum J{\"u}lich, 52425 J{\"u}lich, Germany}

\address{$^2$Quantum Research Centre, Technology Innovation Institute, Abu Dhabi, UAE}

\address{$^3$Pico group, QTF Centre of Excellence, Department of Applied Physics, Aalto University, 00076 Aalto, Finland}

\address{$^4$Moscow Institute of Physics and Technology, 141700 Dolgoprudny, Russia}

\ead{g.catelani@fz-juelich.de}

\begin{abstract}
The fundamental excitations in superconductors -- Bogoliubov quasi\-particles -- can be either a resource or a liability in superconducting devices: they are what enables photon detection in microwave kinetic inductance detectors, but they are a source of errors in qubits and electron pumps. To improve operation of the latter devices, ways to mitigate quasiparticle effects have been devised; in particular, combining different materials quasiparticles can be trapped where they do no harm and their generation can be impeded. We review recent developments in these mitigation efforts and discuss open questions. 
\end{abstract}


\section{Introduction}

Quasiparticle excitations in superconductors have been under investigation, in a sense, since the earliest days of superconductivity: they constitute the ``normal fluid'' part of the phenomenological two-fluid model, the superfluid part being the condensate of Cooper pairs. The nature of these excitations became clear with the advent of BCS theory, according to which the quasiparticles are a coherent superpositions of electrons and holes. This is expressed mathematically by the Bogoliubov transformation
\begin{equation}\label{eq:Bogoliubov}
    \gamma_{k+} = u_k c_{k\uparrow} - v_k c^\dagger_{-k\downarrow} \, , \qquad 
    \gamma_{k-} = u_k c_{-k\downarrow} + v_k c^\dagger_{k\uparrow}
\end{equation}
which can be used to diagonalize the mean-field BCS Hamiltonian, obtain the self-consistent equation for the superconducting gap $\Delta$, and relate it to the critical temperature $T_c$ for the transition to the superconducting state, $\Delta \simeq 1.76 k_\mathrm{B} T_c$. Here $\gamma_{k\sigma}$ denotes the quasiparticle creation operator with momentum $k$ and pseudospin $\sigma=\pm$, $c_{k\uparrow}$ the electron annihilation operator for spin up, and $u_k$ and $v_k$ are the so-called coherence factors, which are functions of the single-electron energies and the gap, see Ref.~\cite{Tinkham} for details.

Quasiparticles are responsible, among other effects, for finite ac losses in superconductors; these losses were first understood in terms of linear response in thermal equilibrium~\cite{MB}. At temperatures well below the critical one (typically down to few tens of mK in aluminum devices with $T_c\gtrsim 1.2\,$K), no equilibrium quasiparticles should be present, so these losses can in principle be avoided.
However, quasiparticles can be driven out of equilibrium, which not only alters the response~\cite{nagaev,devisser}, but can affect superconductivity itself: one of the most striking manifestations of non-equilibrium is the enhancement of the critical temperature with the application of microwaves: in aluminum films, an increase of the critical temperature $T_c$ by about 1\% was observed in Ref.~\cite{tcenh}. 
A photon with energy much higher than $2\Delta$ can break a Cooper pair into two quasiparticles of high energy; these decay by emitting phonons that can break additional pairs into more quasiparticles of lower energy, thus generating a large number of quasiparticles. Detecting such a cascade is the working principle of kinetic inductance detectors~\cite{KID} and superconducting nanowires single-photon detectors~\cite{SNSPD}. Here, in contrast, we focus on situations in which quasiparticles are responsible for unwanted effects, in particular in superconducting qubits, microrefrigerators, and electron pumps and turnstiles.\footnote{Even in kinetic inductance detectors, having too many quasiparticles can limit the detector sensitivity, due to generation-recombination noise~\cite{kidgrn}.}

\section{Quasiparticles as sources of errors}

Superconducting circuits comprising Josephson junctions were proposed over 20 years ago as a possible platform for qubits~\cite{shnirman}, and experimental observation of coherent behavior was reported soon after~\cite{bouchiat,nakamura} in a Cooper pair box, a qubit design in which the charging energy $E_C$ is much larger than the Josephson energy $E_J = \hbar I_c/2e$, with $I_c$ the junction critical current. The detrimental effect of quasiparticles on coherence and the need to devise strategies for its limitation were experimentally recognized early on~\cite{lang,aumentado}, and backaction from the measurement was identified as a source of quasiparticle poisoning~\cite{lukens}. Theoretical works initially focused on the Cooper pair box~\cite{lutchyn1,lutchyn2}, and the relevance of nonequilibrium quasiparticles was further highlighted in Ref.~\cite{martinis}. A systematic treatment of quasiparticle effects for arbitrary qubit designs was developed in a series of works~\cite{catelani1,catelani2,catelani3} by treating perturbatively (as appropriate for tunnel junctions) the tunneling Hamiltonian $H_T$
that describes the tunneling of an electron from one side of a junction to the other side. In the quasiparticle picture, this Hamiltonian can be separated into two terms, $H_T = H_T^\mathrm{qp} + H_T^\mathrm{p}$; in both terms, the tunneling amplitude depends on the phase difference across the junction, thus coupling  the qubit degree of freedom to quasiparticles. The term $H_T^\mathrm{qp}$ accounts for single-quasiparticle tunneling events; during such an event a quasiparticle can absorb energy from the qubit, causing its decay. The term $H_T^\mathrm{p}$ describes the coherent tunneling of pairs that lead to the Josephson energy $E_J$. The perturbative treatment, recently reviewed in Ref.~\cite{leshouches}, predicts not only the qubit decay rate due to quasiparticles, but also a change in its frequency. Quantitative comparison between theory and experiments was performed by quasiparticle injection at low temperature in a phase qubit~\cite{martinis2} as well as by measurements as function of temperature in a transmon qubit~\cite{paik}, two designs with $E_J \gg E_C$: for the decay rate, its proportionality to the quasiparticle density was thus verified.\footnote{Qubit decay due to thermal equilibrium quasiparticles was, in retrospect, already measured in Ref.~\cite{lisenfeld}, but tentatively attributed there to two-level systems.} The quasiparticle density normalized by the density of Cooper pairs, often denoted as $x_\mathrm{qp}$, is a useful measure of deviation from thermal equilibrium: for $T/T_c = 0.01$, a value routinely reached in experiments, the thermal equilibrium expectation is $x_\mathrm{qp} \simeq 10^{-77}$, while typical experimental estimates range between $10^{-9}$ and $10^{-5}$. In the flux-tunable fluxonium~\cite{manucharyan}, the interference between the electron and the hole parts of the quasiparticles [cf. Eq.~(\ref{eq:Bogoliubov})] leads to an enhancement of the decay time when half a flux quantum threads the qubit loop~\cite{pop}.

In addition to tunnel-junction based superconducting qubits, other approaches have been proposed, such as the Andreev level qubit~\cite{ALth}, in which a subgap level formed at a phase-biased, highly-transparent junction can be occupied by zero, one, or two quasiparticles. Coherent superposition between the zero and two quasiparticle states has been demonstrated, but poisoning by a single quasiparticle disrupts the qubit~\cite{ALex}. Similarly, such poisoning affects proposed qubits based on proximitized nanowires hosting Majorana states~\cite{rainis}. To keep the presentation compact, we will not discuss further Andreev and Majorana states in this paper.

Superconducting~\cite{geerligs} and normal-superconductor hybrid~\cite{averin_pekola08} devices have been proposed as current sources for metrological applications, see~\cite{pekola_rmp} for a review. Either a single Cooper pair or a single electron is transferred through the device at a rate directly proportional to the frequency $f$ at which certain gate voltages are changed, leading to a current $I = q f$, with $q=2e$ for Cooper pair pumps and $q=e$ for single-electron pumps.
In both types of devices, quasiparticles can limit the accuracy in transferring precisely one charge in each operation cycle, and thus cause deviation from the expected current quantization. The loss of accuracy can be understood by a simple overheating picture. Take as an example the hybrid turnstile~\cite{averin_pekola08,pekola_rmp}, where electrons from a normal metal island of a single-electron transistor are pumped into a superconductor, there forming quasiparticle excitations at energies $\sim \Delta$. Then in cyclic operation the power to the superconducting lead is $P \sim \Delta\cdot f$. The injection of heat is balanced by diffusion along the superconductor and electron-phonon relaxation, both 
processes weakening at low temperatures, as we explain in the next section.
All these heat fluxes together determine the steady state temperature of the superconducting electrode right at the junction, and due to the weak relaxation, overheating is inevitable: without extra precautions, effective temperatures in the range of 300 to 500~mK for an aluminum lead with $T_c\approx 1.3\,$K are quite common. Such dramatic overheating increases the number of quasiparticles and hence degrades the pumping accuracy substantially. The situation is even more critical in biased fully superconducting structures, where heat diffusion in islands isolated by tunnel junctions is essentially prohibited.

Microrefigerators based on NIS junctions were proposed and demonstrated over 20 years ago~\cite{nahum,leivo,muhonen}. The cooling of the N electrode takes place when the junction is biased at a voltage $V\sim \pm\Delta/e$ as a result of skimming the electrons off from N with energies above the Fermi level -- a form of evaporative cooling. Similarly to the hybrid turnstile, tunneling electrons create quasiparticle excitations with energy $\sim \Delta$ into S and power $\sim I \Delta/e$ is injected into it, where $I$ is the current through the junction. Again the S electrode overheats, and backtunneling of hot quasiparticles decreases the cooling power and efficiency of the refrigerator.

As summarized above, quasiparticles can adversely impact the performance of various superconducting and hybrid devices. Therefore the ability to control the dynamics and generation of quasiparticles can be beneficial, as we discuss in the next sections.

\section{Controlling quasiparticle dynamics}

\begin{figure}[bt]
    \centering
    \includegraphics[width=0.9\textwidth]{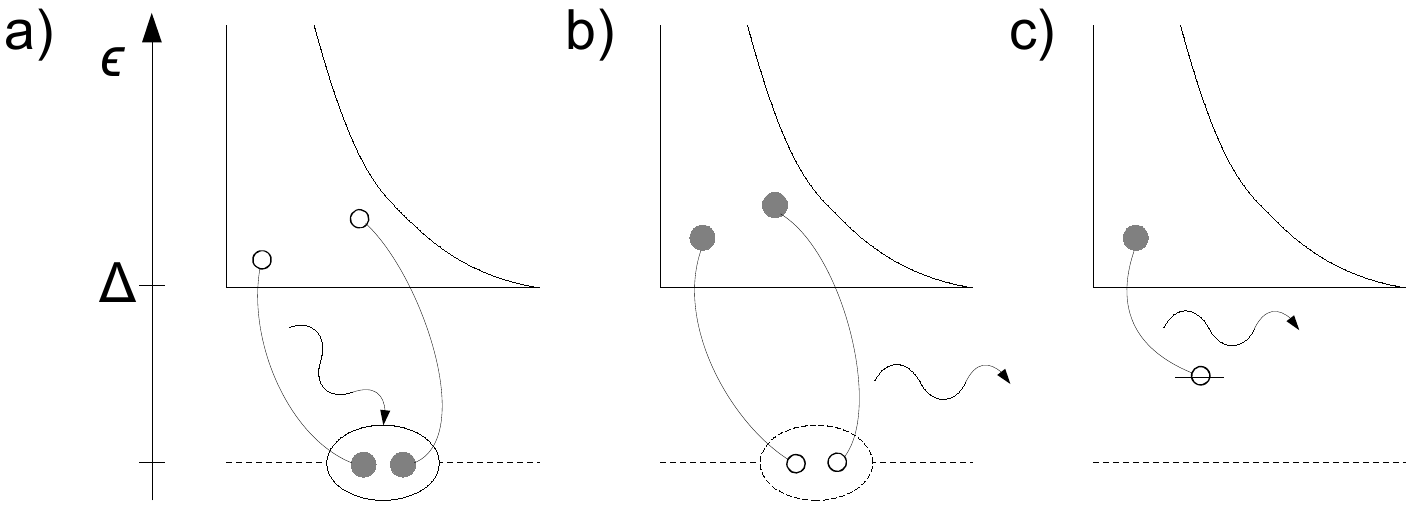}
    \caption{Schematic representation of quasiparticle generation, recombination, and trapping. The horizontal dashed lines give the position of the Fermi energy, with the superconducting density of state (solid curves) starting at energy $\Delta$ above that. a) Generation takes place when a photon or phonon (arrow) of energy larger than twice the gap $\Delta$ breaks a Cooper pair. b) two quasiparticles can recombine into a Cooper pair by emitting a phonon. c) a quasiparticle can be trapped in a state with energy below the gap; such a state can be in the core of a vortex, in a normal or (lower gap) superconducting trap, or a localized state in the superconductor itself (see also Fig.~\ref{fig:mit}).}
    \label{fig:dyn}
\end{figure}

Quasiparticle dynamics is governed by various mechanisms: generation, recombination, diffusion, and trapping -- see Fig.~\ref{fig:dyn}. We postpone the discussion of generation to the next section, focusing here on the other mechanisms. 

Recombination is the mechanism by which two quasiparticles join to form a Cooper pair which then returns to the condensate. This process is accompanied by the emission of a phonon with energy larger than $2\Delta$, so the recombination rate is related to the strength of the  electron-phonon interaction. As recombination requires two excitations to meet, its rate is expected to be exponentially suppressed at low temperatures due to the expected small number of quasiparticles, $f_\mathrm{qp}(\epsilon) \sim e^{-\epsilon/k_\mathrm{B}T}$ at $k_B T \ll \Delta$. However, experimental evidence for a saturation below $T/T_c \sim 0.2$ in the recombination-limited quasiparticle relaxation time~\cite{barends2008,timofeev} points to the possibility that either a different (not electron-phonon) mechanism becomes relevant, or that the electron-phonon coupling is affected by disorder at those low temperatures. In fact disorder, more likely at the surface, appears to play a role in determining the value of this relaxation time~\cite{barends2009}, which in aluminum is of order $100\,\mu$s. A recombination time of this magnitude was also measured for exactly two quasiparticles in a small aluminum island~\cite{maisi}. For thin superconducting films, properties of the substrate and of the interface between the two materials are also important: resonators fabricated on membranes show slower relaxation back to the steady state as compared to samples in which the membrane is in contact with a much thicker wafer~\cite{vercruyssen}, and acoustic mismatch between the superconducting material and the substrate lead to an effective recombination time longer than in the bulk~\cite{kaplan}, since phonons emitted during recombination can break a Cooper pair before escaping to the substrate.

Quasiparticle diffusion is similar to that of electrons in disordered metals, the main difference being that their diffusion constant $D_\mathrm{qp}$ depends on their energy $\epsilon$, ideally tending to zero approaching the gap, $D_\mathrm{qp} \to 0$ as $\epsilon \to \Delta$, while approaching the normal-state value well above it, $D_\mathrm{QP} \simeq D_\mathrm{N}$ for $\epsilon \gg \Delta$; when averaging over the quasiparticle distribution function $f_\mathrm{qp}(\epsilon)$, the diffusion constant remains finite but depends for example on temperature, as verified experimentally~\cite{ullom1998}. Together with the low-temperature exponential suppression of the number of quasiparticle, this leads to an ineffective heat transport, the electronic part $\kappa_\mathrm{S}$ of the thermal conductivity being $\kappa_\mathrm{S}/\kappa_\mathrm{N} \sim e^{-\Delta/k_\mathrm{B}T}$ with $\kappa_\mathrm{N}$ the thermal conductivity in the normal state. These consideration show that at low temperatures it is inefficient to rely solely on diffusion to remove excess quasiparticles or heat from superconducting parts of devices; thus the interest in modifying the dynamics.

\begin{figure}[bt]
    \centering
    \includegraphics[width=0.9\textwidth]{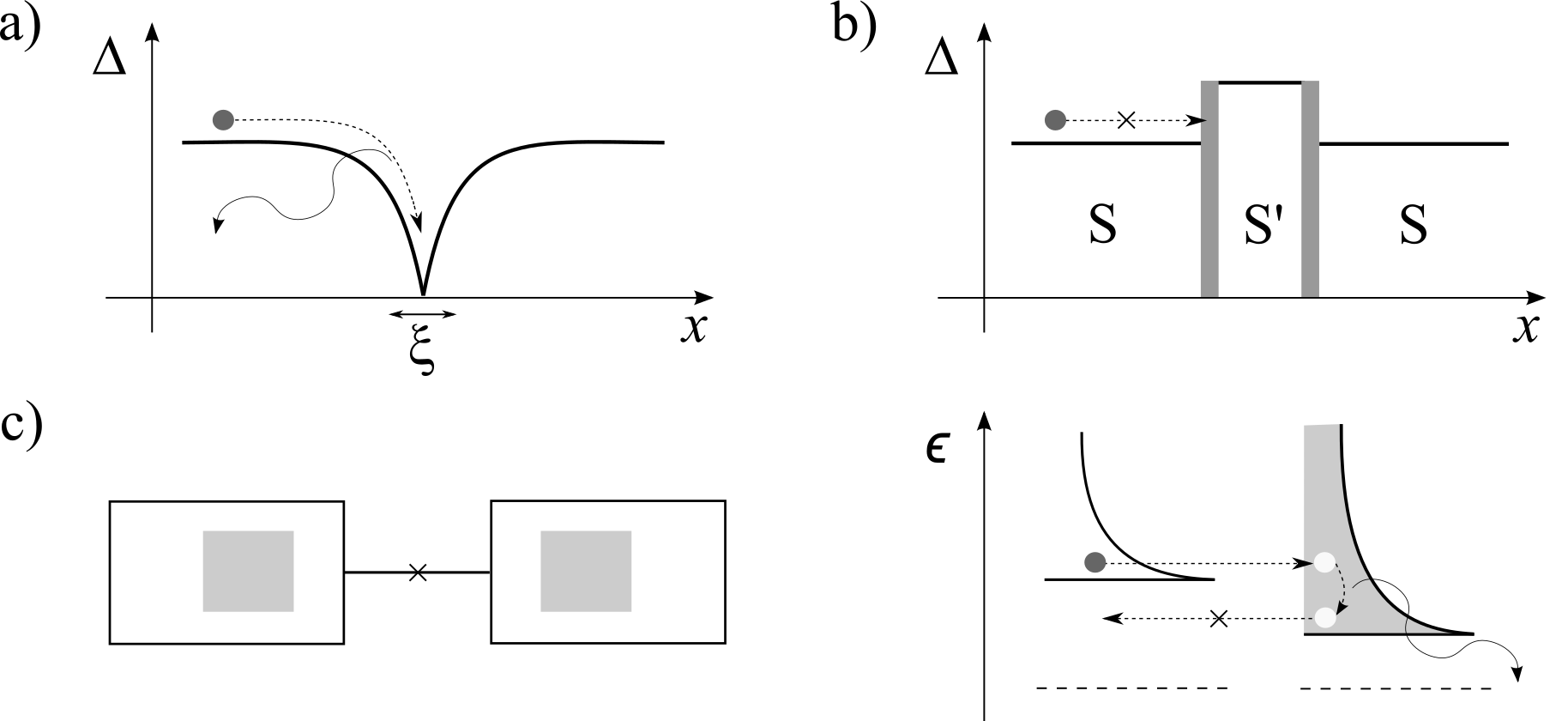}
    \caption{Sketch of various quasiparticle mitigation strategies. a) In the presence of a vortex, the order parameter $\Delta$ (solid line) is suppressed over a core region of size comparable to the coherence length $\xi$; a quasiparticle can be trapped in the vortex core if it loses energy, e.g. by phonon emission.  b) Example of gap engineering: a superconducting island (S') has gap higher than the leads (S); a low-energy quasiparticle cannot tunnel into the island. c) Left: schematic of a transmon qubit with two superconducting pads connected by a Josephson junction (cross); an island of lower gap superconducting material (gray) is overlaid on each pad. Right: a quasiparticle that goes from the higher to the lower gap superconductor and then loses energy cannot tunnel back into the pad.}
    \label{fig:mit}
\end{figure}

Early evidence that vortices can affect the quasiparticle dynamics dates to experiments in the 1990's with superconducting detectors whose performance degrades in the presence of a magnetic field, see~\cite{ullom1998b} and references therein. On the contrary, small magnetic fields can improve the cooling power of NIS junctions~\cite{peltonen2011}. In both cases, the effect can be explained by vortices trapping quasiparticles, see Fig.~\ref{fig:mit}a): the detector performance worsens as less quasiparticles are available to be detected, while the cooling power increases because the quasiparticle cannot backtunnel into the normal electrode and their thermalization is faster. Recent works with superconducting resonators~\cite{nsanzineza}, qubits~\cite{wang}, and turnstiles~\cite{taupin} have focused on characterizing the effectiveness of each single vortex in trapping quasiparticles in their cores, which one can qualitatively think as normal-state regions with radius of the order of the coherence length $\xi$. For the same reasons as in coolers, application of magnetic fields is beneficial to turnstile accuracy. The picture for qubits is more complicated: on one hand, vortices can prevent quasiparticles from reaching tunnel junctions and therefore causing qubit decay; on the other hand, the normal vortex core is a source of loss itself~\cite{ku}.

Vortices are not the only way to keep quasiparticles away from sensitive parts of devices: any spatial modulation of the gap has the potential to confine quasiparticles; designing such modulations is generally known as ``gap engineering'', see Fig.~\ref{fig:mit}b). Small superconducting island were predicted~\cite{av_naz} and observed~\cite{tuominen,lafarge} to display parity effects, since an additional, unpaired electron has a free energy cost equaling the gap at $T=0$. However, such parity effects can be masked by the presence of excess quasiparticles: this has been verified by studying gap-engineered Cooper pair transistors in which the central island can have higher or lower gap than the leads to which it is connected; the difference in gap can be achieved by changing oxygen doping in Al ~\cite{aumentado}, or by using different materials such as Ti/Al~\cite{macleod} and V/Al~\cite{shimada}. An island (NbTiN) with higher gap than the leads (Al) can preserve its parity for a long time~\cite{vanw}, although this does not mean that quasiparticles are not generated in the meantime~\cite{mannila19}, as we will discuss in the next section. Moreover, the higher gap does not necessarily prevent the island from being poisoned by quasiparticles: for instance, material properties of Nb, likely causing subgap states to be present, can counteract the expected protection from the higher gap~\cite{savin}.


As an alternative to directly modulating the gap within the main components of a device, one can add so-called traps, whose goal is again that of reducing the number of quasiparticles in sensitive regions, or to enable the excitations to efficiently release their excess energy to the phonon bath. Normal-metal traps can be viewed as an extreme form of gap engineering, as the gap in the normal parts is absent. Such traps have been implemented by depositing metals such as Au and Cu so that they are in contact with the superconducting electrodes of NIS refrigerators~\cite{ullom2012} and SINIS single-electron transistors~\cite{saira} and turnstiles~\cite{knowles}; the traps indicated possible improvement in cooling and current quantization accuracy. Subsequent experimental works used instead Mn-doped Al as the normal metal and explored different geometries of the electrodes~\cite{nguyen,peltonen2017}. Traps made of Cu have also been added to transmon qubits~\cite{riwar1} and their optimal design investigated~\cite{hosseinkhani1}. A possible drawback of normal traps is that they induce subgap states by the inverse proximity effect; such states can adversely affect all junction-based devices, but their impact can be minimized by placing the traps a few coherence lengths $\xi$ away from the junctions~\cite{hosseinkhani2,wilhelm}. An additional issue for qubits is the Ohmic losses in the normal metal, which can lead to qubit relaxation; the latter can be minimized with appropriate trap design~\cite{riwar2}, or greatly suppressed by using superconducting traps instead, whose gap is lower than that in the rest of the qubit and that are in good contact with the superconductor forming the qubit, see Fig.~\ref{fig:mit}c)~\cite{riwar3}. Another advantage of superconducting traps is that they can be smaller in size compared to normal traps of the same effectiveness, although they should still be few to several microns long. In contrast to such a ``passive'' SS' configuration, SIS' traps can be actively controlled (and used also as quasiparticle injectors) via a voltage bias, as demonstrated in a superconducting SET~\cite{ferguson} and in a SINIS turnstile~\cite{suarez}. Similarly, in qubits comprising island too small to host traps, an active quasiparticle pumping scheme has been shown to lower the average quasiparticle number~\cite{gustavsson}, and in an approach complementary to trapping, a region with high gap an be used to keep quasiparticles away from a low-gap region~\cite{menard}.
 
In concluding this section, we note that in disordered superconductors there is evidence of traps related to the intrinsic inhomogeneity of the gap; this non-engineered type of traps has been proposed as a possible cause of excess quasiparticles, since spatial confinement that keeps quasiparticles separated from each other impedes recombination~\cite{bespalov}. The untrapping of localized quasiparticle by photons has been invoked to explain the increase of resonator quality factors with photon numbers~\cite{grun}, and it has been suggested that the localized quasiparticle could mimic the effects of two-level systems~\cite{degraaf}.

\section{Controlling quasiparticle generation}

In the preceding section we have given an overview of mitigation strategies for excess quasiparticles that are ubiquitous in superconducting devices. A potentially more effective strategy would be to limit the number of excess quasiparticles to begin with. To eliminate the sources of quasiparticle, one first has to identify them. We can broadly classify these sources in two categories, either as due to back-action from other parts of the experiment, or due to extrinsic causes. Moreover, we can separately discuss generation originating from pair-breaking photons or phonons.

As an example of back-action, let us consider Cooper pair transistors used as electrometers. In Ref.~\cite{naaman} it was shown that the operation of such devices is accompanied by the emission of radiation, although at usual working points the photons' frequencies are below the pair-breaking value $2\Delta/h$. Potentially more dangerous is the emission by detectors of phonons with energy above $2\Delta$. Such phonons have been shown to be generated by the recombination of quasiparticles in NIS injectors and to affect resonators over mm-size chips~\cite{patel}; a similar process is at work in single flux quantum devices used to control superconducting qubits~\cite{leonard}. The emission of pair-breaking phonons by Cooper pair transistors was confirmed in Ref.~\cite{mannila21a}.

Turning to extrinsic sources, in Ref.~\cite{houzet} the absorption of pair-breaking photons in a single-junction transmon, an instance of photon-assisted tunneling, has been shown to explain experimental data for transition rates in such a qubit that would otherwise require quasiparticles to have a high effective temperature~\cite{serniak}, in contrast to what is expected theoretically~\cite{basko}. Summing the contributions of pair-breaking photons and thermally excited quasiparticles can explain the temperature dependence of the transition rates~\cite{leshouches}, and improved high-frequency filtering can suppress them~\cite{serniak2}. For applications such as SINIS turnstiles, capacitive shunting of the junctions to a conductive ground plane is also effective in suppressing photon-assisted tunneling~\cite{pekola2010}.

Phonons due to extrinsic sources are with high probability generated in the substrate of the devices: the substrate has a much bigger volume than the metallic elements deposited on it and hence much higher probability to absorb energy from high-energy particles passing through it.  Since phonons quickly spread throughout the whole chip, they can lead to correlated and hence difficult to correct errors in quantum computing systems; this is why recently a number of works have focused on identifying the sources of phonons (such as muons from cosmic rays and environmental radioactive decays) and quantifying their effects on qubits~\cite{vepsalainen,cardani,martinis2020,wilen,mcewen,gordon}. Intriguingly, a mesoscopic superconducting island (in a chip comprising a ground plane) has been shown to be completely free of quasiparticles -- meaning no pair-breaking events -- for seconds, with a pair breaking rate decreasing with time after cool-down~\cite{mannila21b}; however, this rate is higher than that measured in the  experiments just mentioned, which leaves open the question of its cause.
The propagation of phonons in the substrate was detected early on using arrays of kinetic inductance detectors~\cite{swenson}. In order to mitigate the effects of these phonons, islands of lower-gap superconducting material have been successfully implemented as ``phonon traps'' to reduce quasiparticle generation in resonators~\cite{henriques}.

\section{Summary and outlook}

In this paper we have given an overview of the efforts aimed at controlling quasiparticles, focusing in particular on the developments over approximately the last decade. While we have considered here mostly devices based on tunnel junctions such as superconducting qubits, pumps, turnstiles, and microrefigerators, the findings and approaches described can be useful in improving other devices, ranging from Andreev and Majorana qubits to kinetic inductance and superconducting nanowire detectors. 

Devices for quantum information processing work typically near equilibrium, since ideally the only excitations present are those needed to encode qubit states. In contrast, operation of many other superconducting devices takes place far from equilibrium. In the first case one needs to make sure that there are as few as possible excess quasiparticles generated by external, often poorly controlled or even fully uncontrollable sources. In the latter case devices are biased by sources that generate chemical potential drops across contacts in the structure. Examples of these non-equilibrium devices discussed in this article are electron turnstiles and refrigerators. In these systems useful functionality is achieved by the current $I$ induced due to these chemical potential drops, leading to inevitable quasiparticle generation rate of $I/e$ in a superconductor. The mitigation strategies in the case of these non-equilibrium devices are similar as in those operating ideally near equilibrium. Yet the operation is usually more tolerant to minute quasiparticle densities, the typical condition being a small normalized density $x_\mathrm{qp} \ll 1$.


All aspects of quasiparticle dynamics, namely generation, recombination, trapping, and diffusion, are influenced by the properties of the superconducting materials employed, their surfaces, and interfaces with other materials in the device, such as substrates and engineered traps. However, the majority of devices are based on Al as the superconductor and Cu as the normal metal. As the current devices require further improvements,  there are clear opportunities. One direction is to tune the material properties and fabrication processes; for instance, it is well known that resistivity and gap of Al can be changed by modifying the oxygen pressure during deposition, a process that also modifies the quasiparticle dynamics~\cite{valenti}. Recently it has been shown that reduced variations in junction parameters can be obtained via an optimized fabrication process~\cite{Kreikebaum}, or via laser annealing~\cite{hertzberg}.  In qubits, different materials and designs have been explored to suppress decoherence by two-level systems~\cite{murray}, but the impact on quasiparticles has not been fully assessed yet, as two-level systems are in many cases the dominant source of relaxation. In light of the recent findings about the impact of radioactivity on quasiparticle generation, a complementary direction is to search for optimal combinations of surface treatments, and superconducting, substrate, and traps materials and designs; as an example, a phononic crystal substrate has been shown to slow down recombination~\cite{rostem}, and such an approach could perhaps be adapted to hinder generation.

\ack

We thank R.-P. Riwar and H. Hsu for their critical reading of the manuscript.

\end{document}